\documentclass{PoS}

\usepackage{lineno}

\title{Transverse single-spin asymmetry of weak
bosons and Drell-Yan production in p+p collisions at STAR: present and future}

\ShortTitle{Kinematically reconstructed Weak Bosons at STAR}

\author{\speaker{Salvatore Fazio}\thanks{For the STAR Collaboration}\\
        Brookhaven National Laboratory\\
        E-mail: \email{sfazio@bnl.gov}}

\abstract{Accessing the Sivers TMD function in proton+proton collisions through the measurement of transverse single
spin asymmetries (TSSAs) in Drell-Yan and weak boson production is an effective path to test the fundamental
QCD prediction of the non-universality of the Sivers function. Furthermore, it provides 
data to study the spin-flavor structure of valence and sea quarks inside the proton and to test the evolution of parton distributions. 
The TSSA amplitude, $A_{N}$, has been measured at STAR in proton+proton collisions at 
$\sqrt{s} = 500$~GeV, with a recorded integrated luminosity of 25~pb$^{-1}$. 
Within relatively large statistical uncertainties, the current data favor theoretical models that include a change of sign for the Sivers function relative to observations 
in SIDIS measurements, if TMD evolution effects are small. 
RHIC plans to run proton+proton collisions of transversely polarized beams at $\sqrt{s} = 510$~GeV in 2017, delivering an integrated luminosity of 400~pb$^{-1}$. This will allow STAR to perform a precise measurement of TSSAs in both Drell-Yan and weak boson production. The present status and future plans for the Sivers function program at STAR will be discussed as well as other observables sensitive to the non-universality of the Sivers function in the Twist-3 framework, e.g. the TSSA of direct photons.
}

\FullConference{XXIV International Workshop on Deep-Inelastic Scattering and Related Subjects\\
		11-15 April, 2016\\
		DESY Hamburg, Germany}

\begin{document}

\section{Introduction}
Understanding the partonic structure of the proton in multi-dimensions has become a hot topic in the past decade~\cite{Chen:2015}. 
Transversely polarized spin effects can be used to access transverse momentum dependent (TMD)~\cite{Aybat:2011} parton distribution functions (PDFs), 
which contain information on the intrinsic transverse momentum of a parton, together with the fraction that the same parton carries of the longitudinal momentum of the parent nucleon, leading to a 2+1 dimensional picture of the proton. 
Drell-Yan and $W^{\pm}$/$Z^{0}$ boson production in p-p collisions are observables which provide the two
scales typically required to apply the TMD framework to transverse single-spin asymmetries. 
One hard scale is given by the invariant mass, while a soft scale is given by the transverse momentum.
A particularly interesting TMD is the so-called Sivers function~\cite{Sivers:1991}, $f^{\perp}_{1T}$, which describes the correlation of parton transverse momentum with the transverse spin of the nucleon.

There is evidence of a quark Sivers effect in semi-inclusive DIS (SIDIS) measurements~\cite{Anselmino:2009} where the quark Sivers function is associated with a final state effect from the gluon exchange between the struck quark and the target nucleon remnants. On the other hand, for the $DY$ process or the $W^{\pm}/Z^{0}$ production in p-p collisions, the Sivers asymmetry originates from the initial state of the interaction. As a consequence, the different color flow leads to the fundamental prediction that the quark Sivers functions are of opposite sign in SIDIS and in $DY/W^{\pm}/Z^{0}$~\cite{Collins:2002}
\begin{equation}
f^{SIDIS}_{1T^\perp} (x, k_\perp) = - f^{DY/W^{\pm}/Z^{0}}_{1T^\perp} (x, k_\perp),
\end{equation}
 and this non-universality is a fundamental prediction from the gauge invariance of QCD. 

The experimental test of this sign change is one of the open questions in hadronic physics, and can provide insights on the TMD factorization~\cite{Collins:2002}. While luminosity and experimental requirements for a meaningful measurement of asymmetries in Drell-Yan production are challenging, weak boson production is also sensitive~\cite{Echevarria:2014} to the predicted sign change and can be measured at STAR.  

Thanks to the high $Q^{2}\simeq M^{2}_{W^{\pm}/Z^{0}}$ scale, the weak boson production provides also a stringent test of the TMD evolution~\cite{Echevarria:2014}. Furthermore, the $W^{+}(W^{-})$ boson, which is produced through $u+\bar{d} (d+\bar{u})$ annihilation, can provide essential input to disentangle 
the contribution to the Sivers function of light-sea quarks, which is essentially unconstrained by fits to SIDIS data~\cite{Echevarria:2014}.
The STAR experiment at RHIC is currently the only place in the world where all these effects can be tested simultaneously.

The transverse single-spin asymmetry, $A_N$, 
solely calculated from the decay lepton is a very strong function of the lepton kinematics~\cite{Kang:2009bp} and therefore its measurement is experimentally challenging. Therfore a full reconstruction of the produced boson kinematics is crucial for a meaningful measurement.

\section{Recent results}

Based on the transversely polarized data sample corresponding to a luminosity of $L = 25$ pb$^{-1}$ collected in the year 2011 at $\sqrt{s}=500$~GeV, 
the STAR Collabortion has recently published~\cite{STAR:An_WeakBosons} the world's first measurement of $A_N$ in weak boson production. 
The $W^{\pm}$ boson momentum has been fully reconstructed from the decay lepton and all other particles in the recoil from the initial hard scattering.

Background contributions coming from $W^{\pm}\rightarrow \tau^{\pm} \nu_{\tau}$, $Z^{0} \rightarrow e^{+}e^{-}$ have been studied using PYTHIA 6.4~\cite{PYTHIA6.4}, whereas background from QCD events has been studied using a data driven procedure reversing our signed-$P_{T}$-balance selection cut. All background sources have been estimated to be at most a few percent of the selected sample. 

\begin{figure}[htbp]
  \centering
  \includegraphics[width=1.0\textwidth]{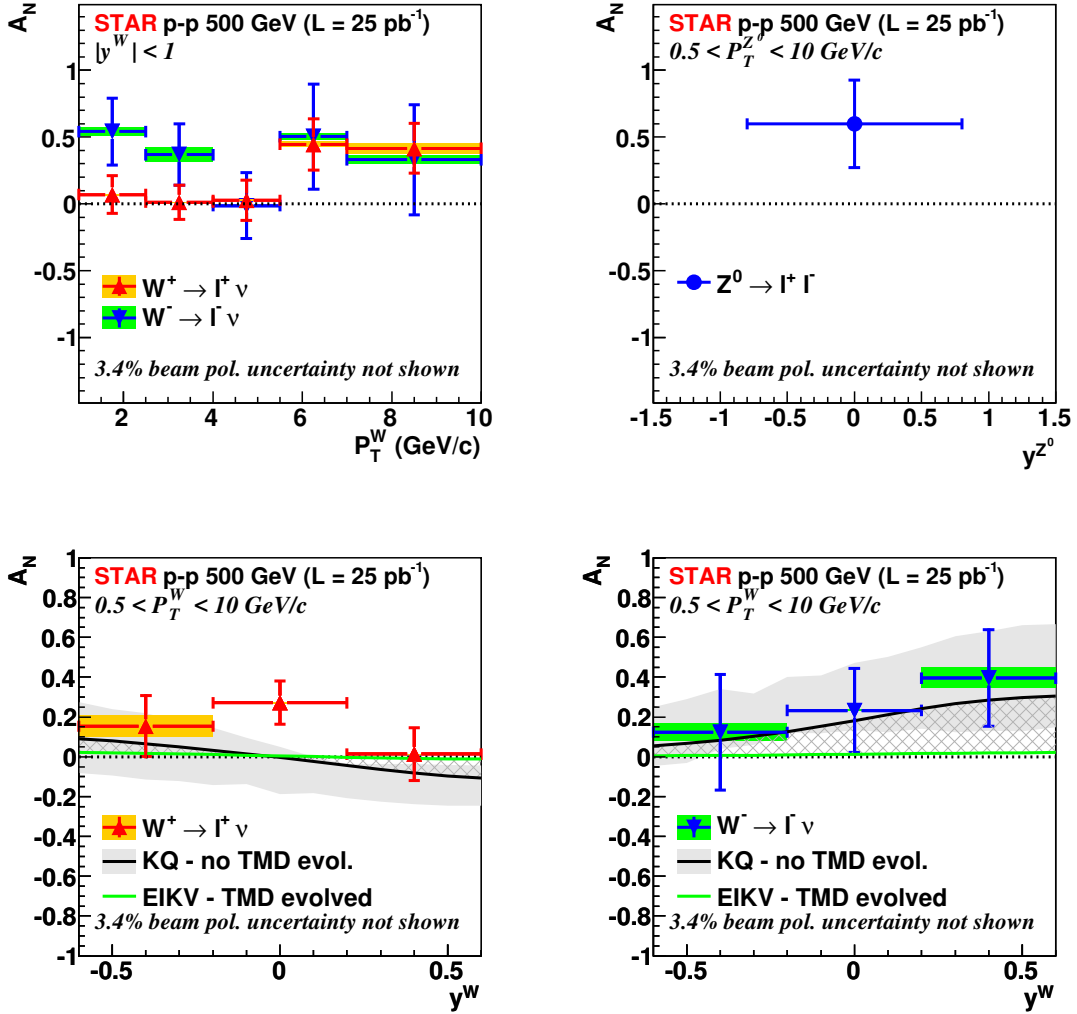}
  \caption{Transverse single-spin asymmetry amplitude for $W^{\pm}$ and $Z^{0}$ boson production measured at STAR in a pilot run at $\sqrt{s}=500$~GeV with a recorded luminosity of 25~pb$^{-1}$.}
  \label{Fig:W-An}
\end{figure}

The novel STAR results for the $A_{N}$ measurement of the $W^{+}$ and $W^{-}$ boson production are shown separately in Fig.~\ref{Fig:W-An} as a function of $y^{W}$ and $P_{T}^{W}$. 
The systematic uncertainties, added in quadrature, have been evaluated via a Monte Carlo test using a theoretical prediction for the asymmetry from~\cite{Echevarria:2014}. The 3.4\% overall systematic uncertainty on beam polarization measurement is not shown in the plots.

The measured amplitudes are compared with theoretical predictions from KQ~\cite{Kang:2009bp}, which does not account for TMD evolution, and from EIKV~\cite{Echevarria:2014}, which depicts the largest evolution effects among all current theoretical calculations. The gray band in Fig.~\ref{Fig:W-An} highlights the uncertainty on the EIKV calculation due to the current ignorance of the sea-quark Sivers function.
Despite the large statistical uncertainties, global fits to the current data favor theoretical models that include a change of sign for the Sivers function relative to observations
in SIDIS measurements, if TMD evolution effects are small.

The transverse single-spin asymmetry of the $Z^{0}\rightarrow e^{+}e^{-}$ process has been also measured because it has  many advantages:  it is experimentally very clean and the boson kinematics are easily reconstructed 
from the two decay leptons produced at central rapidities, well within the acceptance of the STAR detector. Thus, the measurement is background free, carries only the overall systematic uncertainty coming from the polarization measurement, and the asymmetry is expected to be of the same size as that of the $W^{\pm}$. The only challenge comes from the much lower cross section of the $Z^{0}\rightarrow e^{+}e^{-}$ process, which requires the collection of a very large data sample for a statistically significant measurement. 

The STAR result for the $A_{N}$ measurement of the $Z^{0}$ boson production in a single $y^{Z}$, $P_{T}^{Z}$ bin is shown in Fig.~\ref{Fig:W-An}. 

\section{Future program}
This recently published analysis demonstrates the capability of STAR to perform measurements with fully reconstructed $W^{\pm}$ bosons.  
A larger data sample is needed for a conclusive test of the change in the sign of
the Sivers function, to decompose the contribution from sea-quarks and to pin down the TMD evolution.

\begin{figure}[htbp]
  \centering
  \includegraphics[width=0.65\textwidth]{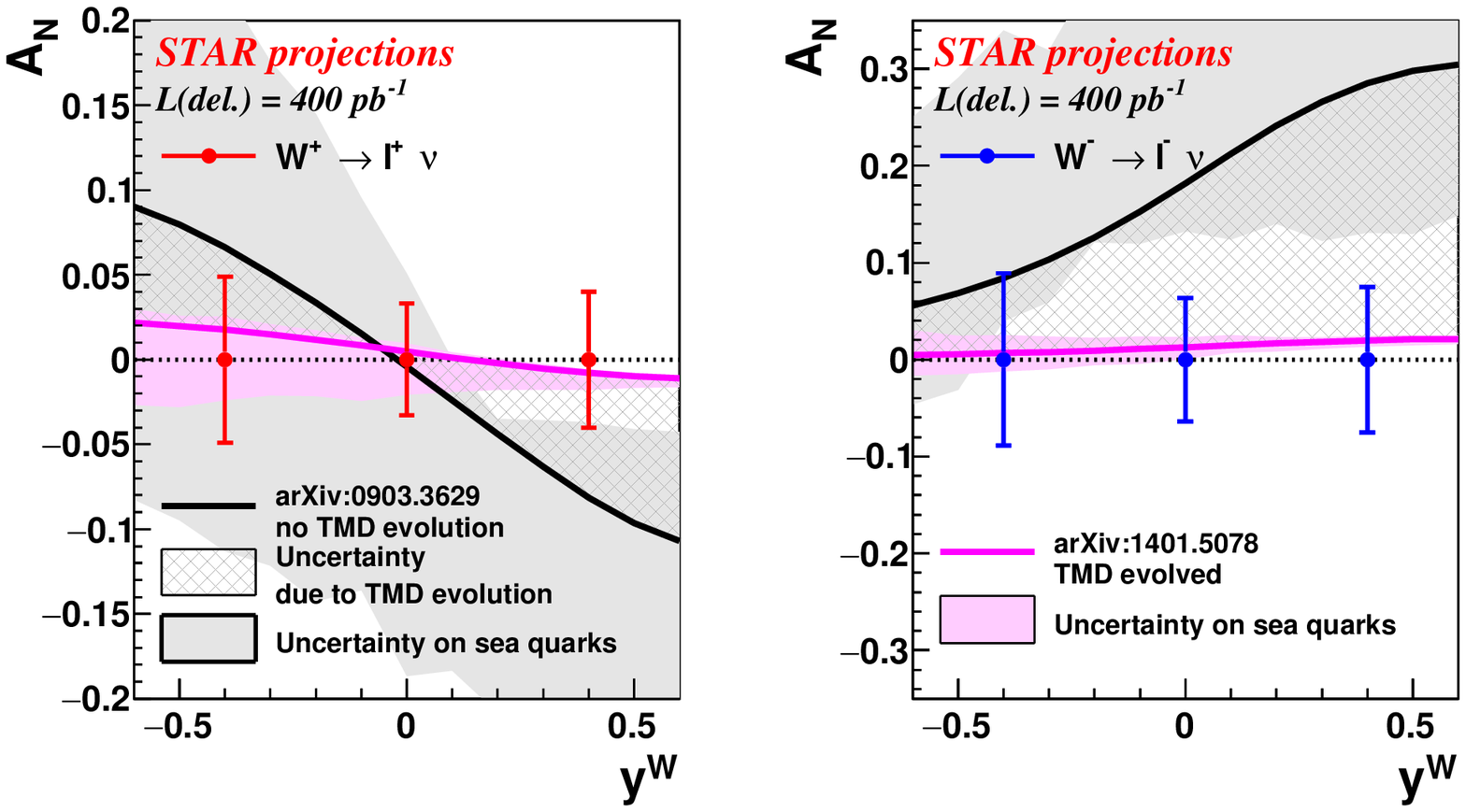}
  \includegraphics[width=0.34\textwidth]{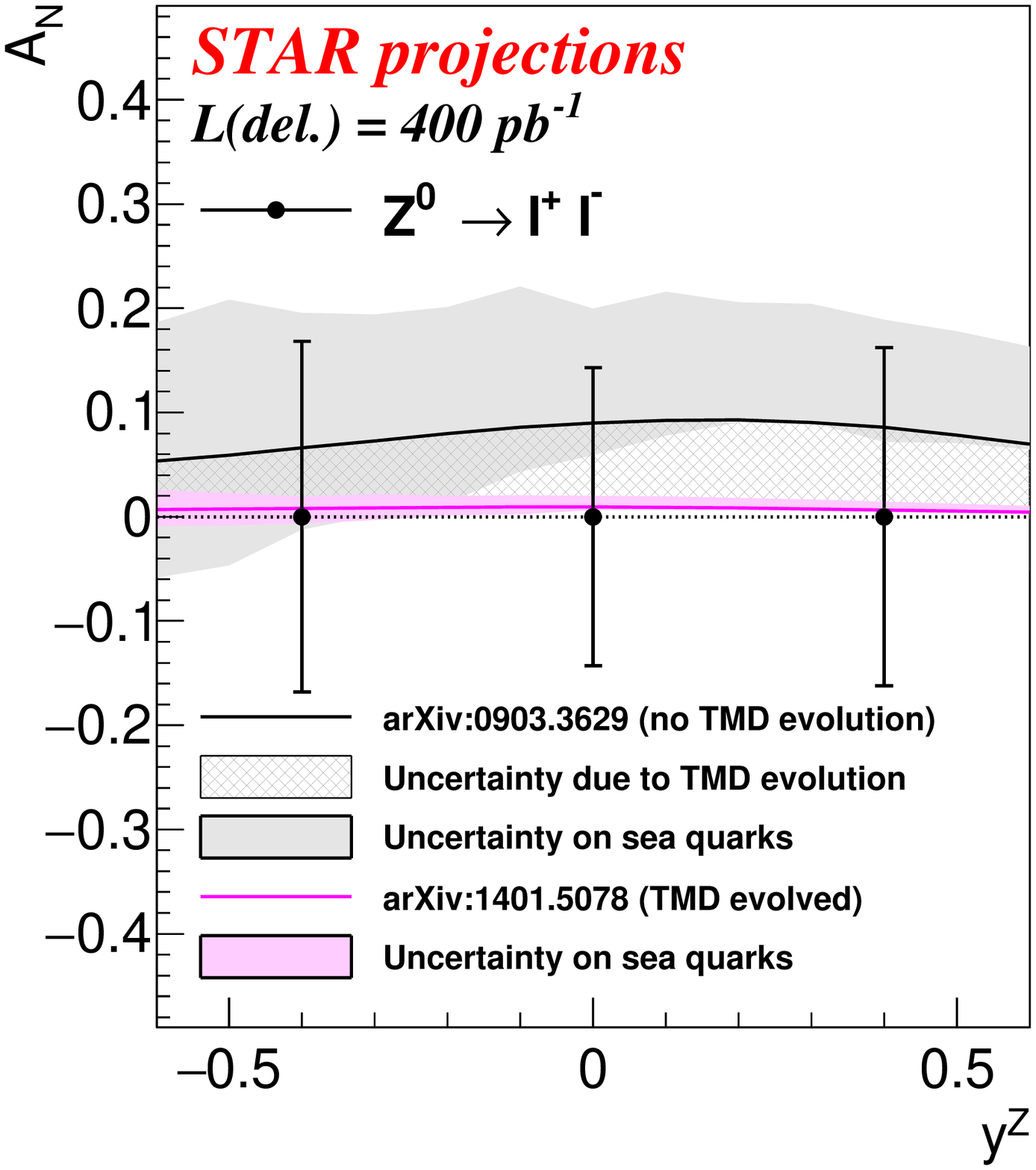}
  \caption{The projected uncertainties for the $A_{N}$ of W$^{\pm}$ and Z$^{0}$ as a function of the boson
rapidity for a delivered integrated luminosity of 400~pb$^{-1}$, and an average beam polarization of 55\%.}
  \label{Fig:WZ-projections}
\end{figure}

RHIC plans to collect 400~pb$^{-1}$ events of transversely polarized p+p collisions at $\sqrt{s}=510$~GeV during the 2017 run, using a dynamic $\beta$* squeeze~\cite{beta_squeeze} throughout the fill which will reduce the pileup effects in the tracker and increase the boson selection efficiency. 
This will allow for a precise $A_{N}$ measurement of weak boson production at STAR (for projections see Fig.~\ref{Fig:WZ-projections}), and can lead to the first experimental test of the sign change of the Sivers function if the evolution effects on the $A_{N}$ prove to be smaller than a factor $\sim 5$. 
Furthermore it will provide an ideal tool to study the spin-flavor structure of sea quarks inside the proton.

STAR will be also pursuing the investigation of TMDs and the experimental test of the Sivers sign change through measuring $A_{N}$ in Drell-Yan. This measurement is particularly challenging as it requires a severe suppression of the overwhelming hadronic background, which is of the order of $10^{5} \sim 10^{6}$ larger than the signal. In order to achieve the electron/hadron discrimination, necessary for such background suppression, STAR envisions to install a post-shower detector at the forward rapidity range $2.5 < \eta < 4.0$ behind the Forward Pre-shower Spectrometer (FPS) and the Forward Meson Spectrometer (FMS), to create a three-detector system~\cite{ColdQCD}.

\begin{figure}[htbp]
  \centering
  \includegraphics[width=0.55\textwidth]{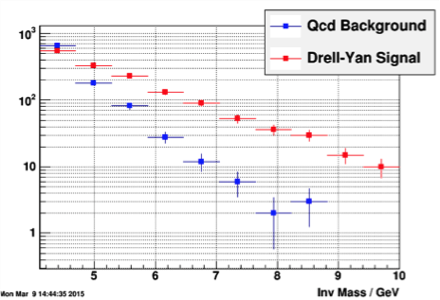}
  \includegraphics[width=0.4\textwidth]{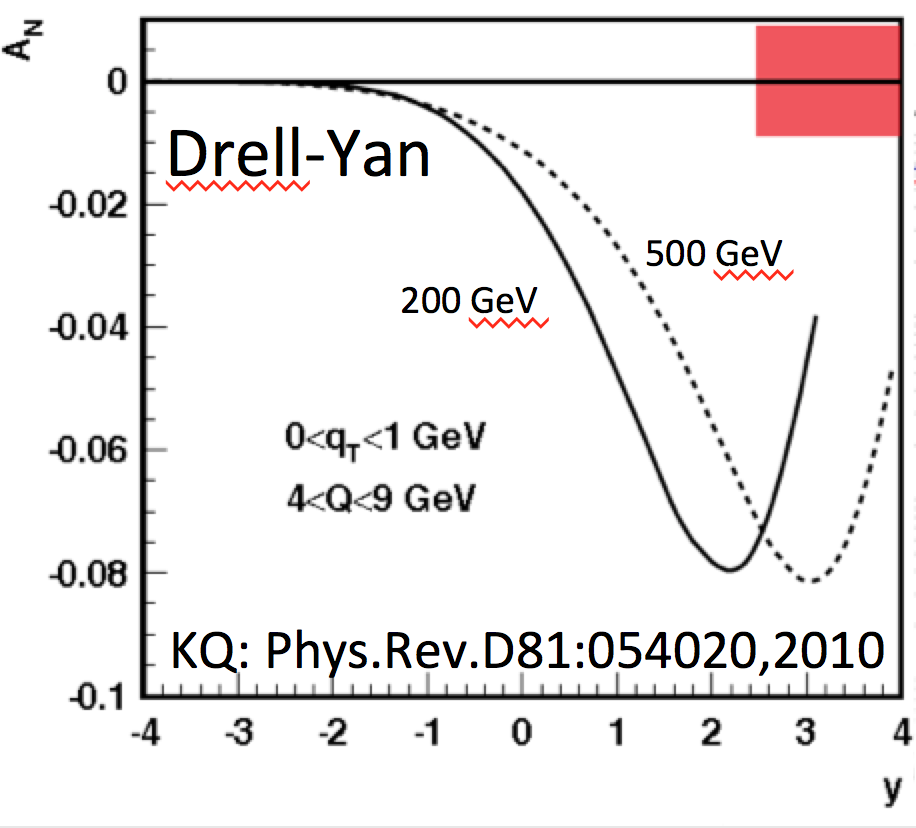}
  \caption{Left: Drell-Yan signal (red) and QCD background (blue) as a function of the invariant mass. Right: the projected uncertainties for the $A_{N}$ in Drell-Yan in the corresponding rapidity-bin (red area) for a delivered integrated luminosity of 400~pb$^{-1}$.}
  \label{Fig:DY-plots}
\end{figure}

Monte Carlo simulation was used to estimate signal-to-background ratio for Drell-Yan to QCD events. 
For the background simulation, all basic QCD $2 \rightarrow 2$ processes as well as heavy-flavor channels
were generated, and events were smeared using the simulated detector resolutions of the forward detector
system. The Drell-Yan production through quark-antiquark annihilation and quark-gluon scattering processes
were separately generated and scaled to the same luminosity. The final yields of Drell-Yan events (signal)
and background events are shown in Fig.~\ref{Fig:DY-plots} ({\it left}). Clearly, the proposed forward-detector system upgrade provides the needed rejection factor to allow for a measurement of the TSSA in Drell-Yan. The projected uncertainty is shown in Fig.~\ref{Fig:DY-plots} ({\it right}). 

\begin{figure}[htbp]
  \centering
  \includegraphics[width=0.6\textwidth]{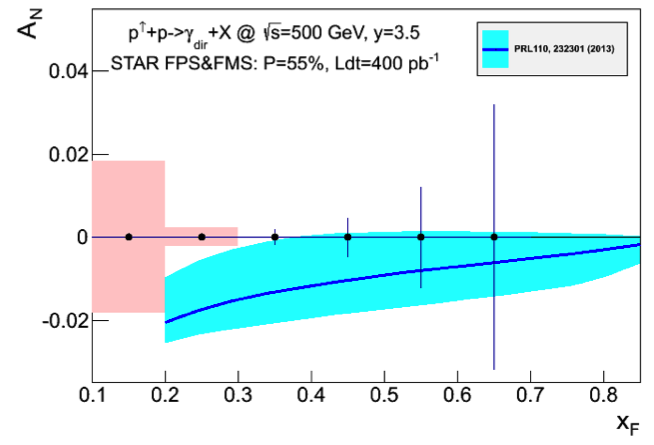}  
  \caption{Statistical  and  systematic  uncertainties  for  the  direct  photon  $A_{N}$ compared  to  theoretical  prediction~\cite{Gamerg:2013} for a delivered integrated luminosity of 400~pb$^{-1}$, and an average beam polarization of 55\%.}
  \label{Fig:photons}
\end{figure}

Using the same data collected during the 2017 RHIC run, STAR plans also to investigate the Sivers function sign change indirectly by measuring single scale processes like $A_{N}$ for prompt photons. 
These observables can be described in the Twist-3 formalism but the relevant functions appearing in this framework can be related to the Sivers function.
The assumption is that if the correlation due to different color interactions for initial and final states between the Sivers function and the Twist-3 function 
would be violated, the asymmetry would be positive but of the same magnitude.
The projected uncertainties for the $A_{N}$ measurement in direct  photon, after  background  subtraction, is shown in Fig.~\ref{Fig:photons} and compared  to  theoretical  predictions~\cite{Gamerg:2013}.

The STAR experiment at RHIC is currently the only experiment capable of investigating the TMD evolution effects, studying the contribution from sea-quarks to the Sivers function, and providing an ultimate experimental test of the non-universality of the Sivers function.

This can be done 
by measuring the TSSA for a variety of processes - W$^{\pm}$, Z$^{0}$, Dell-Yan, prompt-$\gamma$ - simultaneously at the same experiment.

\end{document}